# Observation of Electromagnetic Transients in a Nb$_3$Sn 4-layer Dipole Mirror Magnet

Steven Krave, Maria Baldini, Igor Novitski

*Abstract*—During testing of the Stress Managed Cosine-Theta dipole mirror magnet SMCTM1, magnet quenches were observed following large voltage spikes in the half-coil voltage taps which preceded normal quench initiation. Following some recent work at CERN measuring power current converter transients in relation to flux jumps, an additional differential current probe was added to the magnet instrumentation. Measurements of this probe, as well as the half coil voltage taps show a fast and substantial current change preceding the quench, hinting at coil motion. A simple model based on relative shift of the inner and outer layer coils was developed which suggests the magnitude of coil shift events is on the estimated relative coil shift from the magnet constraint conditions. The behavior is shown to be reversible and characteristic of stick-slip.

## I. INTRODUCTION

SMCTM1 (Stress Managed Cosine-Theta Magnet 1) is a 4-layer Nb$_3$Sn stress managed cosine-theta dipole mirror magnet designed and built at Fermilab within the scope of the U.S. Magnet Development Program (US-MDP) [1]. It consists of two double layer cosine-theta coils with the inner coil being of a more traditional design and the outer coil SMCT1 being stress-managed where the cable is wound into a 3D printed stainless steel structure that consists of a spar element that is integrated with the coil wedges forming a single piece, as opposed to standard wedges that are normally free-floating in a magnet cross section. To manage the complex geometry of the outer coil support structure, it was fabricated by additive manufacturing technology using a selective laser melting process [2].

A section view of the magnet structure can be seen in **Fig. 1**. With the inner coil nested inside of the stress managed coil with minimal radial gap. Radial prestress is applied via a hydraulic press and aluminum clamps that are inserted via hydraulic cylinder. On cool down, this pre-stress is maintained through differential contraction of the aluminum clamps and iron yoke.

After magnet assembly, it received the standard instrumentation of strain gages, voltage taps, and acoustic sensors. An image of the completed mirror magnet before testing, highlighting the location of the acoustic sensors can be seen in Fig. 2.

The SMCT1 coil test in a dipole mirror structure was done in two configurations - SMCTM1a with powering the SMCT1 (layers 3-4) coil only, and SMCTM1b with powering the SMCT1 coil in series with the MDP03 (layers 1-2) coil. The objective of the tests was to prove the SMCT coil concept in 2-layer and 4-layer mirror configurations; demonstrate that the magnet can reach the target current (coil field) at the established preload; study magnet training, training memory after thermal cycle, ramp rate and temperature dependences of the magnet quench current. The quench performance of this magnet can be found in [3].

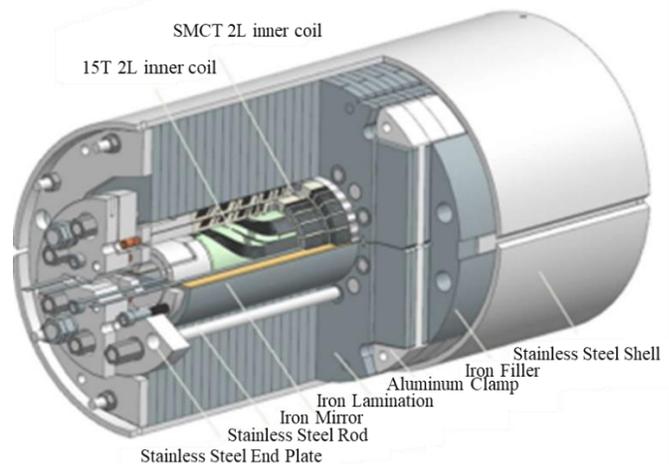

**Fig. 1.** 3D view of the 4-layer dipole mirror magnet SMCTM1. Note that each coil end is supported independently

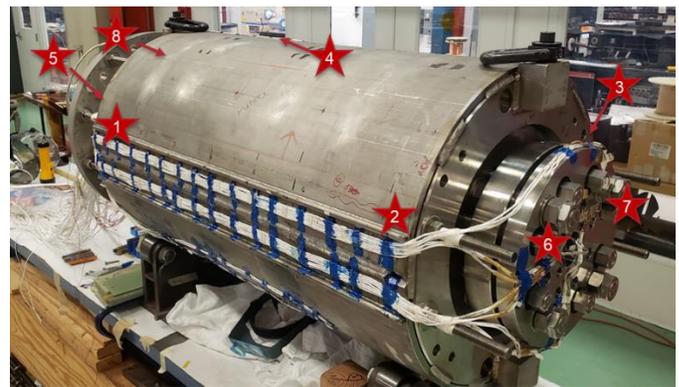

**Fig. 2.** Assembled magnet and location of acoustic sensors.

### A. Voltage spikes during testing

During testing the initial magnet training in the 4-layer

This work was produced by FermiForward Discovery Group, LLC under Contract No. 89243024CSC000002 with the U.S. Department of Energy, Office of Science, Office of High Energy Physics. Publisher acknowledges the

U.S. Government license to provide public access under the DOE Public Access Plan DOE Public Access Plan

Color versions of one or more of the figures in this article are available online at http://ieeexplore.ieee.org

configuration, the magnet protection system was experiencing nuisance trips from voltage spikes that were not corresponding to a voltage growth typically associated with a magnet quench [3]. This behavior was remedied by the installation of a 1$^{st}$ order lowpass filter to smooth (reduce) the spike behavior at the FPGA input and slightly delay system response. This minor modification allowed magnet training to proceed as usual.

With the trip issue managed it was observed that several of the training quenches were initiated with a voltage spike event that developed into a quench. This voltage spike appeared to correspond to an acoustic event, which appeared like other events throughout the ramp.

On closer investigation of the acoustic events associated with the voltage spikes, it became apparent that the acoustic sensors were picking up some sort of electrical noise at each event and not sensitive to acoustics. For reference a typical acoustic event can be seen in Fig. 3. In contrast to the expected acoustic signatures, the sensor data were highly asymmetric and with a fast decay indicating the electrical issues as in Fig. 4. In some sense the sensors were acting as uncharacterized electrical pickups. The sensor malfunction was traced to a polarity mismatch of the interface cable used for sensor checkout.

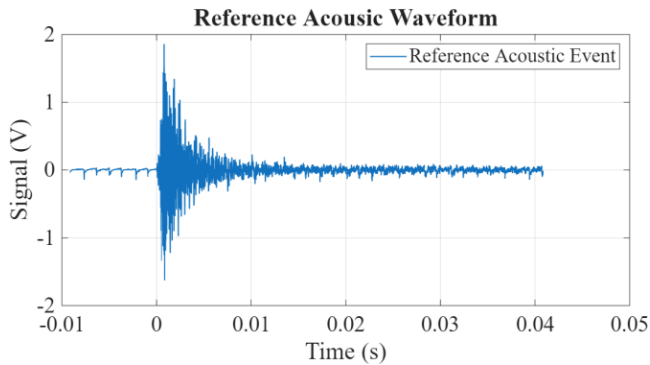

**Fig. 3.** Sample Acoustic data from typical acoustic event. Note reasonably symmetric waveform and long decay. This response is typical for all sensors.

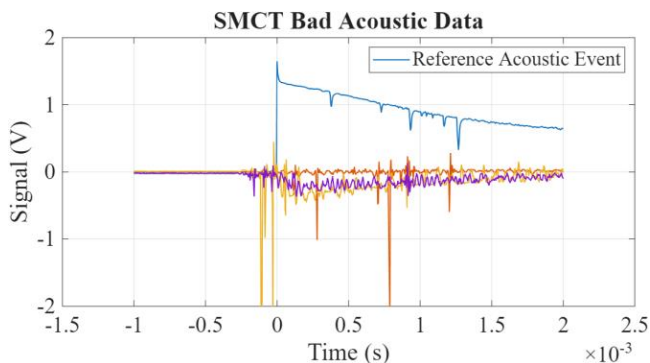

**Fig. 4.** Acoustic sensor 'data' during ramping. CH A-D are malfunctioning acoustic sensors on either end of the magnet with unusual response to events.

The voltage spikes seen in the half-coil signal and acoustic electrical noise were concurrent, which suggests some sort of large electromagnetic transients may be occurring. Following the work done at CERN on correlation of voltage spikes with small changes in operating current, a similar investigation was undertaken [4].

The integrated current measurement in the Fermilab Vertical Magnet Test Facility (VMTF) was not sensitive enough for the expected mA level current variation on top of the ~10 kA operating current so an additional commercial clamp on Rogowski coil was added over the test facility leads as seen in Fig. 5. This configuration essentially rejects the DC background component and is only sensitive to the small ac transients present on the facility leads. With a sensitivity of around 1 mA with some filtering this becomes a 100 ppb level measurement.

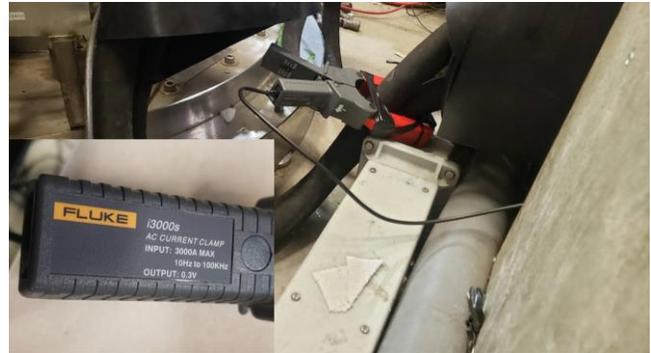

**Fig. 5.** Current Clamp installed on test facility leads.

With the addition of the current clamp, it is clear that the half-coil spikes, 'acoustic' spikes and current transients are all correlated as seen in Fig. 6.

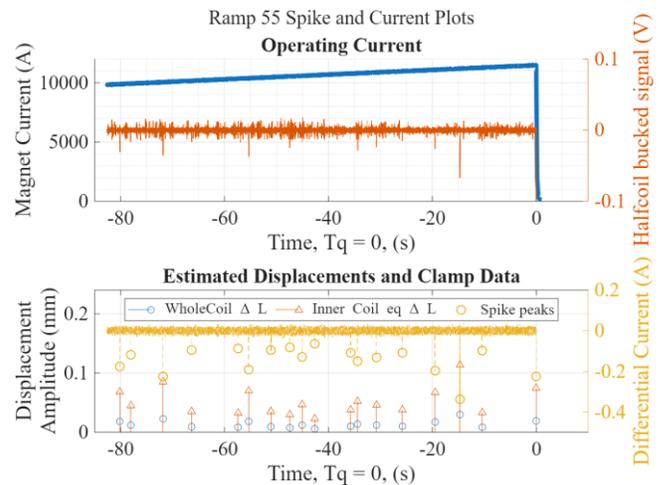

**Fig. 6.** Addition of half-coil bucked voltage signal shows voltage spike associated with combined 'acoustic' data. Channels are shifted and scaled for clarity.

## II. ESTIMATION OF COIL MOTION

With the correlation of the electrical signals to a current transient established, it becomes possible to make an estimate of coil motion based on known or measurable system parameters. As established in [4], conservation of energy can be used to calculate an effective inductance change corresponding to the amplitude of a current transient.

Since this magnet and structure are extremely rigid in the cross section as a result of the integral metal printed structure and solid iron yoke, it is assumed that any coil motion present consists of a stretching or contraction along the magnet axis. In the case of motion along the magnet axis, the inductance is linear with magnet length, aside from the 3d contribution at the ends, and is already known from the magnet design and electrical checkout. To double check this value during operation to account for effective inductance including transients, we cross check vs the di/dt from a magnet extraction and dump resistor. A representation of this measurement is seen in Fig. 7.

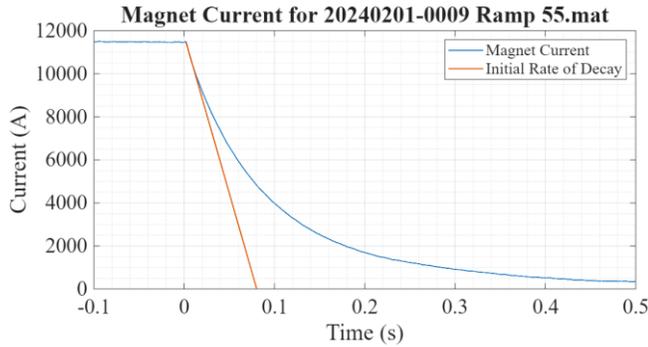

**Fig. 7**. Illustration of calculation of magnet inductance. di/dt is calculated from first few ms of dump.

In the case of this magnet, extraction was completed with a 60 mΩ dump resistor, leading to a calculated overall inductance of 44 mH. This value is in line with the estimated and measured overall magnet inductance. The magnet straight section is 0.5 m for an inductance of 88 mH/m. This corresponds to the full magnet response. Since the two coils are not glued and supported with independent axial support structures, it is reasonable to assume that only the inner coil is slipping with respect to more rigid SMCT coil. In this case the situation becomes more complicated and the difference between inner coil and outer coil is needed as well as the mutual inductance. To be most conservative, we ignore the mutual inductance and look at the relative contribution of the inner coil alone, where the outer coil inductance is 2.8 times larger than the inner, or that the inner would have to move ~3.8 times further than the entire coil assembly for the same change in current. This estimate is partially confirmed from the gain required to balance the quench protection system matches the estimated values from the magnet design. It is understood that these assumptions are somewhat coarse, but they set an upper limit on the overall displacement.

The inductance of the magnet for a spike event is calculated by determining the energy stored in the magnet at the time of the spike at the reference $L_O$ value and equating it to the energy stored at some new current and some new inductance. We can calculate as follows:

$$L_{new} = \frac{L_0 * i_{ref}^2}{(i_{ref} - di_{spike})^2} \quad (1)$$

Where $L_{new}$ is the new calculated inductance of the magnet as a function of the current spike amplitude $di_{spike}$ at the magnet operating current at the time of the event. The difference in inductance before and after the spike can be used with the inductance per meter scalars presented above to calculate a displacement amplitude depending on the assumed motion.

A script to collect events throughout a ramp and calculate the effective displacement for a single coil or whole magnet longitudinal growth was written results are presented below in Fig. 8.

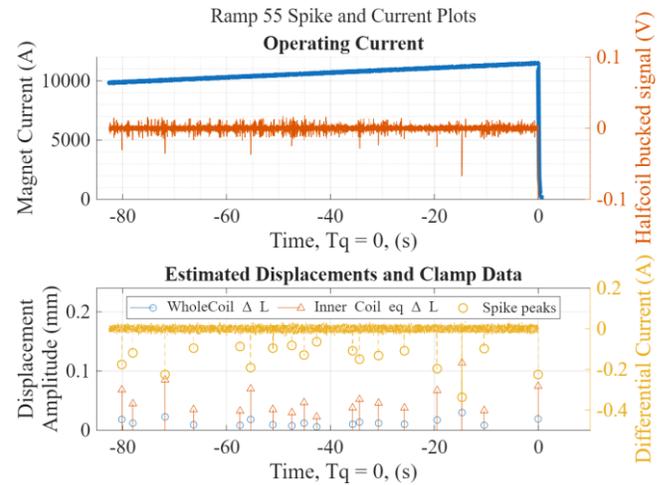

**Fig. 8.** Plot of half-coil data, current clamp data and estimated displacements.

Unfortunately, on this magnet the current clamp was only installed on one ramp with a quench, so the data shown is limited. Still, 16 events were recorded with perfect concurrence to the half-coil signal. Using the above method to calculate the overall coil motion, event amplitudes up to between 230 µm and 850 µm of relative coil motion throughout the recorded period are observed. Ideally this would be cross checked against the strain gages, but the sample rate and resolution of the strain system did not allow this analysis.

Some example cases of these spikes are presented below. Throughout the ramp there are several events like 'Event 15' which occur with similar magnitudes. It is seen that the current spike signal is extremely fast and very prominent. Aside from a 720 Hz comb filter for power supply ripple, no other post processing is done.

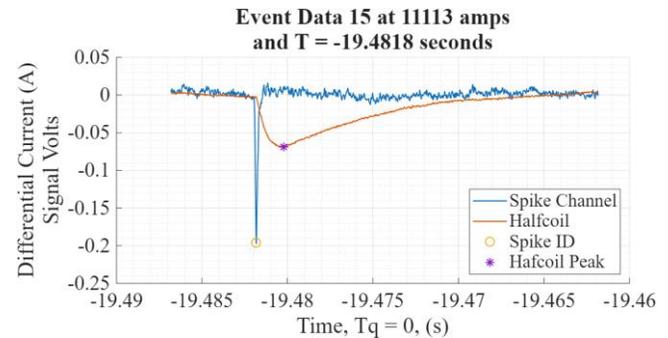

**Fig. 9.** Mid-ramp voltage spike and current transient.

In the case of the mid ramp spike above, the coil voltage tap has some signal which quickly recovers, likely as a result of the system filtering and no normal transition. This behavior is present in all spikes recorded except for the quench event shown in Fig. 10.

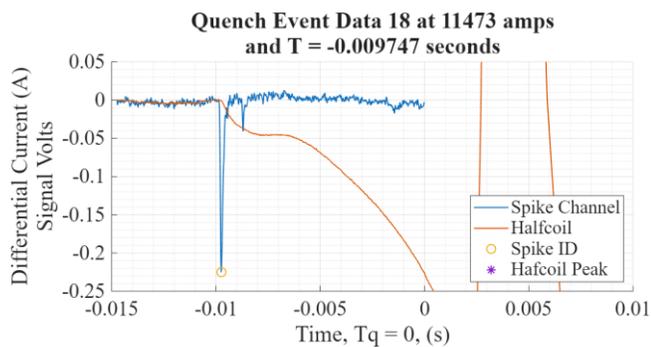

**Fig. 10.** Quench spike event. Current and voltage transient followed by quench leading to voltage runaway.

In this event, the half-coil signal first departs from ~0 to some level of 50 mV from the spike event, which progresses into a resistive growth from initiation of a quench.

*A. Reversibility of Behavior*

Assuming that the previously noted explanation to this behavior is correct, it should be possible to demonstrate that it reverses on ramp down, as the coil and structure relax and spring back to the initial position. In this case, the structure will relax and shrink in length, leading to a positive current and voltage spike contrasted to the negative spikes during ramp up. This behavior can be observed as expected and is shown in Fig. 11. The reverse events do not appear for a substantial portion of ramp down until ~2 kA. In the case of an initial ramp up to 10.5 kA, indicating some hysteresis in the relaxation behavior as expected from the stick-slip aspect of this motion.

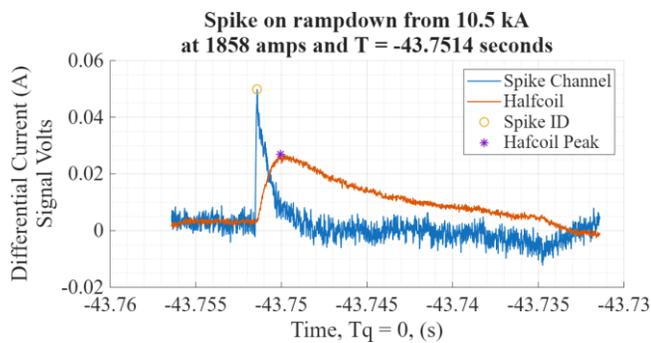

**Fig. 11.** Voltage and current transient during ramp down. Signals are in the opposite direction showing reversed behavior during ramp down.

III. CONCLUSIONS

During the training of SMCT1b, voltage spikes were observed. A Rogowski coil was used to show that these voltage spikes are related to current spikes and are likely related to coil motion, through the same methods used at CERN in several coil tests. This motion would be expected given a frictional constraint between the inner and outer coils resulting in a stick slip expansion/contraction. An attempt is made to calculate the amplitude of this motion which produces reasonable results, though a more precise magnetic model could further refine this information. Based on this analysis, as well as additional mechanical analysis, the dipole mechanical structure is being optimized to improve axial coil support. This measurement has been implemented on recently tested magnets and will be discussed in future work.

IV. ACKNOWLEDGEMENT

We wish to acknowledge the contributions of Sasha Zlobin for his insight on magnet mechanics and Dan Eddy and Alex Yuan for their assistance implementing this measurement.